\documentclass[fleqn,twoside]{article} 
\usepackage{epsf,multicol,ifthen}
\usepackage{ujp}
\usepackage[cp1251]{inputenc}
\usepackage[english,russian]{babel}
\usepackage{amstext}
\usepackage{amssymb}
\usepackage{cite}
\usepackage{graphicx}

\usepackage{bm}
\usepackage{amsopn}
\usepackage{amsmath}
\usepackage{amsthm}

\theoremstyle{plain}

\newtheorem*{proposition*}{╥тхЁфцхээ }

\DeclareMathOperator{\Tr}{Tr} \DeclareMathOperator{\res}{res}
 \DeclareMathOperator{\ad}{ad}
 \DeclareMathOperator{\Mat}{Mat}
\DeclareMathOperator{\ord}{ord} \DeclareMathOperator{\diag}{diag}

\allowdisplaybreaks

\nazva{ORDERED STATES AND NONLINEAR LARGE-SCALE
EXCITATIONS IN A PLANE MAGNET WITH SPIN \boldmath{$s\,{=}\,1$}}%

\udk{538.221}

\nazvacol{ORDERED STATES AND NONLINEAR EXCITATIONS}%

\avtor{J.M. BERNATSKA, P.I. HOLOD}%
\avtorcol{J.M. BERNATSKA, P.I. HOLOD}%

\inst{National University ``Kyiv-Mohyla Academy''}%
\adr{(2, Skovoroda Str., Kyiv 04070, Ukraine), }
\insti{M.M. Bogolyubov Institute for Theoretical Physics, Nat. Acad. of Sci. of Ukraine}%
\adri{(14b, Metrolohichna Str., Kyiv 03143, Ukraine)}%

\begin{document}           
\setcounter{page}{1208}%
\maketitl                 
\begin{multicols}{2}
\anot{%
We study ordered states and topological excitations in a
quasi-two-dimensional magnet, modeled by a square lattice with spins
$s\,{=}\,1$ at all sites, and the Hamiltonian with biquadratic
exchange interaction between nearest neighbor sites. We propose two
effective Hamiltonians for description of large-scale excitations in
the two-dimensional case. They describe excitations of the mean
field in a nematic phase and a mixed ferromagnetic-nematic phase. It
is shown that the effective Hamiltonians are minimized on
configurations with fixed topological charge. These topological
excitations can arise at low temperatures and cause a destruction of
a long-range order in the two-dimensional system.}

\section{Introduction}

Quasi-two-dimensional magnets have various techno\-lo\-gi\-cal
applications. They serve as magnetic films used for recording of
information, thin ferromagnetic layers in Josephson semiconductor
junctions, layered resistive systems, {\it etc.}

Here, we will not deal with applied aspects of theory of magnetism.
However, we note that a study of two-dimensional systems has also a
significant value. Studying ordered states, their stability, and
excitation spectra, we obtain model scenarios of a self-organization
of a substance with decrease in a temperature or under an action of
external fields. To support the above-presented assertion, it is
worth to recall an important role played by the Onsager's results
\cite{Onsager} on the two-dimensional Ising model, or the
Kosterlitz--Thouless theory of topological phase transitions
\cite{Kosterlitz, Berezinsky}. This is also related to the study of
the two-dimensional $\mathrm{O}(3)$-sigma model or a planar
Heisenberg magnet \cite{Tsvelik}.  As a natural continuation of this
trend, we mention a significant number of scientific papers devoted
to investigations of two-dimensional continuous or lattice systems
with high spins at sites.

In the present paper we consider a generalized Heisenberg magnet
taking into account bilinear and biquadratic interactions at
nearest-neighbor sites of a square or cubic lattice. This
Hamiltonian for magnets of spin $s\,{=}\,1$ was proposed and studied
long ago \cite{Blume, Chen} without any restriction on
dimensionality. At the beginning of 1970s, an existence of ordered
phases different from the ferromagnetic or antiferromagnetic ones
was established by the mean-field methods. In particular, if the
constant of biquadratic interaction is larger than that of bilinear
interaction, then a pure quadrupole ordering or a spin nematic state
can be realized in the system \cite{Matveev, Nauciel}.

It is known that a two-dimensional system with a continuous group of
symmetries has no long-range order at $T\,{>}\,0$ (the
Mermin--Wagner theorem). In many cases, an instability of ordered
phases in two-dimensional systems implies an existence of nonlinear
topological excitations caused by small fluctuations of temperature.
The role of such excitations in destruction of a long-range order is
proven within the model of plane rotators \cite{Kosterlitz,
Berezinsky} and for the two-dimensional Heisenberg ferromagnet
\cite{Tsvelik}.

The main result of our work is a proof of an existence of
topological excitations in the model with biquadratic interaction
between nearest spins $s\,{=}\,1$ at sites of a square lattice. We
will consider the boundary case of a nematic phase, where the
constants of bilinear and biquadratic interactions are identical. It
is known that, in this case, energies of both possible phases
(nematic and ferromagnetic-quadrupole ones) are equal. In order to
study excitations of nematic phase, we assume that
$K\,{-}\,J\,{=}\,\varepsilon$, and $\varepsilon$ is a small positive
value. It is obvious that topological excitations exist at these
parameters, and differ slightly from those in the case $\varepsilon
\,{=}\, 0$. If $\varepsilon\,{>}\, 0$, the manifold of degeneration
of a ground state of the system is deformed, but the topology is not
sensitive to smooth deformations. Therefore, our conclusion of
existence of topological excitations at $J\,{=}\,K$ remains valid
also in the case $K\,{>}\,J\,{>}\,0$.

The present paper contains two parts. The first part is a survey. In
the mean field approximation, we reveal existence conditions for
ordered phases and solve an equation of self-consistent relations
for order parameters. Comparing with results of other researchers on
this topic, we obtain conditions of occurrence of a nematic state.
In the second part, averaging equations of motion over coherent
states, and passing from a plane square lattice to a continuous
plane, we obtain formulas for free energy of an inhomogeneous
distribution of the mean field. Topological excitations are
described in terms of the inhomogeneous distribution. Depending on a
choice of an equilibrium state and a way of averaging, we get two
formulas for the free energy: the first formula corresponds to
excitations of a pure nematic state, and the second one is related
to excitations in the state with nonzero magnetization and
quadrupole moment. In the case of $\mathrm{SU}(3)$-symmetry, we
determine self-dual solutions of the problem of minimization. The
obtained topological excitations give the absolute minimum for the
free energy, and its value is proportional to a topological charge.

\section{The Quantum Model of a Planar Magnet}

Let us consider a plane square lattice containing atoms of spin $s$
at each site. Each atom is assigned by three spin operators
$\{\hat{S}^{1}_n,\,\hat{S}^{2}_n,\,\hat{S}^{3}_n\}$ obeying  the
standard commutation relations
\[
[\hat{S}_{n}^{\alpha},\hat{S}_{n}^{\beta}] =
i\varepsilon^{\alpha\beta\gamma} \hat{S}_{n}^{\gamma} \delta_{nm},
\]
where the indices $\alpha$, $\beta$, and $\gamma$ run the values
$\{1,\,2,\,3\}$ for each site $n$, and $\delta_{nm}$ is
the Kronecker symbol.

Usually, such system is described by the Heisenberg Hamiltonian. As
$s\,{\geqslant}\, 1$ we can include higher orders of the exchange
interaction in the Hamiltonian. In particular, magnets with the
biquadratic interaction were studied in the 1970s \cite{Harris,
Rodbell}. The latter Hamiltonian will be considered in what follows.
Let
\begin{equation}\label{BBH}
\hat{\mathcal{H}} = -\sum_{n,\delta}
\{J(\hat{\bm{S}}_n,\hat{\bm{S}}_{n+\delta})+K
(\hat{\bm{S}}_n,\hat{\bm{S}}_{n+\delta})^2\},
\end{equation}
where $\hat{\bm{S}}_n$ denotes the vector
$(\hat{S}_n^1,\,\hat{S}_n^2,\,\hat{S}_n^3)$ of the spin operators at
site $n$, and $\delta$ runs over the nearest-neighbour sites. We
assume that the exchange integrals $J$ and $K$ are positive, i.e. we
consider mainly the ferromagnetic interaction.

The operators $\{\hat{S}^{\alpha}\}$ are defined over the
$(2s\,{+}\,1)$-dimensional space of an irreducible representation of
the group $\mathrm{SU}(2)$. The spin operators generate the complete
matrix algebra over this space (the Burnside theorem). With respect
to the adjoint action $\ad_{\hat{S}^{\alpha}}$, the complete matrix
algebra is divided into a direct sum of irreducible collections of
tensor operators. For example, let us consider the case of
$s\,{=}\,1$. Then for the complete matrix algebra over the
representation space, we have $\dim \Mat_{3\times 3} \,{\simeq}\,
[9] \,{=}\, [1]\,{+}\,[3]\,{+}\,[5]$. Obviously, bases in the three-
and five-dimensional irreducible collections are formed,
respectively, by the operators $\hat{S}^{\alpha}$, and by the tensor
operators of weight 2. The latter are the operators of quadrupole
moment chosen in the form
\[
\hat{Q}_n^{\alpha\beta} = \hat{S}_n^{\alpha} \hat{S}_n^{\beta} +
\hat{S}_n^{\beta} \hat{S}_n^{\alpha},\ \alpha\neq\beta,
\]
\[
\hat{Q}_n^{[2,2]}=(\hat{S}_n^1)^2-(\hat{S}_n^2)^2,
\]
\[
\hat{Q}_n^{[2,0]}=\sqrt{3}\bigl((\hat{S}_n^3)^2-\tfrac{2}{3}\bigr).
\]

A normalization of the operators $\hat{S}^{\alpha}$ is defined by
the relation
\[
(\hat{S}^1)^2 + (\hat{S}^2)^2 + (\hat{S}^3)^2 = s(s+1)\mathbb{I}_3,
\]
which yields $\Tr (\hat{S}^{\alpha})^2 \,{=}\, \tfrac{1}{3}\,
s\,(s\,{+}\,1) (2s\,{+}\,1)$. For $s\,{=}\,1$ we have $\Tr
(\hat{S}^{\alpha})^2 \,{=}\, 2$. We extend such normalization for
all other operators.

Now we  fix the canonical basis $\{|{+}1\rangle,\, |{-}1\rangle,\,
|0\rangle\}$ in the representation space. Then a matrix
representation of the operators of spin and quadrupole moment is as
follows:
\[
 \hat{S}_n^1=\frac{1}{\sqrt{2}}\begin{pmatrix}
 0 & 0 & 1 \\ 0 & 0 & 1 \\ 1 & 1 & 0 \\ \end{pmatrix},\
 \hat{S}_n^2=\frac{1}{\sqrt{2}}\begin{pmatrix}
 0 & 0 & -i \\ 0 & 0 & i \\ i & -i & 0 \\
 \end{pmatrix},
\]
\[
 \hat{S}_n^3=\begin{pmatrix}
 1 & 0 & 0 \\ 0 & -1 & 0 \\ 0 & 0 & 0 \\
 \end{pmatrix}, \quad
  \hat{Q}_n^{[2,0]}=\frac{1}{\sqrt{3}}\begin{pmatrix}
 1 & 0 & 0 \\ 0 & 1 & 0 \\ 0 & 0 & -2 \\
 \end{pmatrix},
\]
\[
 \hat{Q}_n^{12}=\begin{pmatrix}
 0 & -i & 0 \\ i & 0 & 0 \\ 0 & 0 & 0 \\
 \end{pmatrix},\
 \hat{Q}_n^{13}=\frac{1}{\sqrt{2}}\begin{pmatrix}
 0 & 0 & 1 \\ 0 & 0 & -1 \\ 1 & -1 & 0 \\
 \end{pmatrix},
\]
\[
 \hat{Q}_n^{23}=\frac{1}{\sqrt{3}}\begin{pmatrix}
 0 & 0 & -i \\ 0 & 0 & -i \\ i & i & 0 \\
 \end{pmatrix}, \
 \hat{Q}_n^{[2,2]}=\begin{pmatrix}
 0 & 1 & 0 \\ 1 & 0 & 0 \\ 0 & 0 & 0 \\
 \end{pmatrix}.
\]
It is easy to see that the proposed matrices are connected with the
Gell-Mann matrices $\hat{\Lambda}_a$, $a\,{=}\,1$, $2$, $\dots$,
$8$, which also form a basis in $i\mathfrak{su}(3)$. The connection
is given by the linear transformations
\[
  \hat{S}_n^1=\tfrac{1}{\sqrt{2}}(\hat{\Lambda}_4+\hat{\Lambda}_6),\
  \hat{S}_n^2 =\tfrac{1}{\sqrt{2}}(\hat{\Lambda}_5-\hat{\Lambda}_7),
  \ \hat{S}_n^3 = \hat{\Lambda}_3,
\]
\[
  \hat{Q}_n^{12}=\hat{\Lambda}_2,\quad \hat{Q}_n^{[2,0]}=\hat{\Lambda}_8,\quad
  \hat{Q}_n^{[2,2]}=\hat{\Lambda}_1, \vphantom{\tfrac{1}{\sqrt{2}}}
\]
\[
  \hat{Q}_n^{13}=\tfrac{1}{\sqrt{2}}(\hat{\Lambda}_5+\hat{\Lambda}_7),\quad
  \hat{Q}_n^{23}=\tfrac{1}{\sqrt{2}}(\hat{\Lambda}_4-\hat{\Lambda}_6).
\]

By $\{\hat{P}^{a}\}$ we denote the collection of operators
$\{\hat{S}_n^{1}$, $\hat{S}_n^{2}$, $\hat{S}_n^{3}$,
$\hat{Q}_n^{12}$, $\hat{Q}_n^{13}$, $\hat{Q}_n^{23}$,
$\hat{Q}_n^{[2,2]}$, $\hat{Q}_n^{[2,0]}\}$. It is easy to prove that
the commutation relations
\begin{equation*}
 [\hat{P}^{a}_n,\,\hat{P}^{b}_m] = iC_{abc} \hat{P}^{c}_{n}
 \delta_{nm},
\end{equation*}
hold true. Here, the tensor of structure constants $C_{abc}$ is
antisymmetric under a permutation of any pair of indices, and its
nonzero components are
\[
C_{123}=C_{145}=C_{167}=C_{264}=C_{257}=C_{356}=1,
\]
\[
C_{168}= C_{528}=\sqrt{3}, \quad C_{437}=2,
\]
In terms of the operators of spin and quadrupole moment, Hamiltonian
\eqref{BBH} gets the form
\[
    \hat{\mathcal{H}}=-\left(J-\tfrac{1}{2}K\right)\sum_{n,\delta}\sum_{\alpha}
    \hat{S}^{\alpha}_n, \hat{S}^{\alpha}_{n+\delta} -
\]
\begin{equation}\label{BBH_SU2}
    -\tfrac{1}{2}K\sum_{n,\delta}
    \sum_{a}\hat{Q}^{a}_{n} \hat{Q}^{a}_{n+\delta}-\tfrac{4}{3}KN,
\end{equation}
where $N$ denotes the total number of sites of the lattice.
Obviously Hamiltonian remains $\mathrm{SU}(2)$-invariant; hence, the
operators $\hat{S}^{\alpha}_n$ and $\hat{Q}_n^{a}$ are transformed
by formulas of the adjoint representation
\[
\hat{U}\hat{S}_n^{\alpha} \hat{U}^{-1} = \sum_{\beta}
\hat{D}^{\alpha\beta} (\hat{U})\hat{S}_n^{\beta},
\]
\[
\hat{U}\hat{Q}_n^{a} \hat{U}^{-1} = \sum_{b} \hat{D}^{ab}
(\hat{U})\hat{Q}_n^b,\quad \forall \hat{U}\in \mathrm{SU}(2),
\]
where $\hat{D}^{\alpha\beta} (\hat{U})$ and $\hat{D}^{ab} (\hat{U})$
are matrices of the real irreducible representations of
$\mathrm{SU}(2)$ with dimensions 3 and 5, respectively. If
$K\,{=}\,J$, then the $\mathrm{SU}(2)$-symmetry can be extended to
the group $\mathrm{SU}(3)$. In this case, the Hamiltonian
\eqref{BBH_SU2} gets the form
\begin{equation}\label{BBH_SU3}
    \hat{\mathcal{H}}=-\tfrac{1}{2}J \sum_{n,\delta}
    \sum_{a}\hat{P}^{a}_{n} \hat{P}^{a}_{n+\delta}-\tfrac{4}{3}JN.
\end{equation}

To study possible ordered phases of such spin system, we use the
mean-field approximation.

\section{Mean-Field Approximation. Ordered States}
\label{ss:MF}

Now we replace the interaction between spin and quadrupole operators
that is described by Hamiltonian \eqref{BBH_SU2} with an effective
interaction between the operators $\hat{P}^{a}_n$ and the classical
mean field. Components of the mean field are considered proportional
to averages (quasiaverages) of the quantum operators
$\{\hat{P}_n^a\}$. The Hamiltonian in the mean field approximation
has the form
\[
    \hat{\mathcal{H}}_{\text{MF}}=-(J-\tfrac{1}{2}K)\sum_{n,\delta}\sum_{\alpha} \hat{S}_{n}^{\alpha}\langle
    \hat{S}_{n+\delta}^{\alpha}\rangle -
\]
\begin{equation}\label{HamiltMidField}
     - \tfrac{1}{2}K\sum_{n,\delta}\sum_{a} \hat{Q}_{n}^{a}\langle
    \hat{Q}_{n+\delta}^{a}\rangle-\tfrac{4}{3}KN.
\end{equation}
It is worth to give a warning that a direct calculation  of the
averages $\{\langle \hat{S}_{n}^{\alpha}\rangle\}$ and $\{\langle
\hat{Q}_{n}^{a}\rangle\}$, for example by means of the density
matrix $\hat{\rho}(T) \,{=}\, \exp \{-\frac{\mathcal{H}}{kT}\}$,
results in the zero values. This follows from the
$\mathrm{SU}(2)$-symmetry of Hamiltonian \eqref{BBH_SU2}. Nonzero
values of the averages appear if the symmetry is broken. Symmetry
breaking can be stimulated by an external field, which vanishes
after specifying an order in the magnetic system. The quantities
calculated in this way are called ``quasiaverages''
\cite{Bogolyubov}.

Hence, we assume that in our system the nonzero quasiaverages
$\{\langle \hat{S}_{n}^{\alpha}\rangle\}$ and $\{\langle
\hat{Q}_{n}^{a}\rangle\}$ exist, and form a classical 8-component
vector field $\mu_a(x_n)$, $a\,{=}\,1$, $2$, $\dots$, $8$. In order
to obtain nonzero values of $\{\langle \hat{Q}_{n}^{a}\rangle\}$,
the external field must have nonzero gradient. If the mean field is
homogeneous, an action of the group $\mathrm{SU}(2)$ transforms
Hamiltonian \eqref{HamiltMidField} to the form
\[
   \hat{\mathcal{H}}_{\text{MF}}=-(J-\tfrac{1}{2}K)\sum_{n}\hat{S}_{n}^{3}\langle
    \hat{S}^{3}\rangle -
\]
\[
     -\tfrac{1}{2}K\sum_{n}\hat{Q}_{n}^{[2,0]}\langle
    \hat{Q}^{[2,0]}\rangle-\tfrac{4}{3}KN = - \tfrac{4}{3}KN -
\]
\[
     -\sum_{n} \left\{(J-\tfrac{1}{2}K)
    \hat{S}^{3}_n \mu_3+\tfrac{1}{2}K
    \hat{Q}^{[2,0]}_n\mu_8\right\}.
\]

In the case of thermodynamic equilibrium and an infinite lattice,
the fields $\mu_3\,{=}\,\langle \hat{S}^3 \rangle$ and
$\mu_8\,{=}\,\langle \hat{Q}^{[2,0]} \rangle$ are constant, i.e.
have the same values at all points $x_n$ (a homogeneous mean field).
These quantities serve as \emph{order parameters}. Obviously,
$\mu_3$ describes a normalized magnetization (a ratio of
$z$-projection of magnetic moment to a saturation magnetization),
and $\mu_8$ is analogously related to a quadrupole moment.

In the mean field approximation, it is easy to calculate a partition
function for the homogeneous case
\[
NZ(\mu_3,\mu_8,T) = \Tr
e^{-\frac{\mathcal{H}_{{\mathrm{MF}}}}{kT}} = \Tr
e^{-\frac{Nh_{{\mathrm{MF}}}}{kT}},
\]
where $h_{\text{MF}}$ denotes a one-site Hamiltonian
\begin{equation}\label{One-siteHamilt}
h_{\mathrm{MF}} =  -(J-\tfrac{1}{2}K) \mu_3 \hat{S}^3 -
\tfrac{1}{2}K \mu_8 \hat{Q}^{[2,0]}-\tfrac{4}{3}K.
\end{equation}

The introduced mean field makes sense if \emph{self-consistent
relations} are held:
\[
\mu_3 = \langle \hat{S}^3 \rangle_{MF} = \frac{\Tr \hat{S}^3
e^{-\frac{Nh_{\text{MF}}}{kT}}}{\Tr
e^{-\frac{Nh_{\text{MF}}}{kT}}},
\]
\[
 \mu_8 = \langle
\hat{Q}^{[2,0]} \rangle_{MF}= \frac{\Tr \hat{Q}^{[2,0]}
e^{-\frac{Nh_{\text{MF}}}{kT}}}{\Tr e^{-\frac{Nh_{\text{MF}}}{kT}}}.
\]
these relations serve as an analog of the Weiss equation from theory
of ferromagnetism. The averages of operators can be presented via
the partition function:
\[
\mu_3 = \frac{kT}{(J-\frac{K}{2})}\frac{\partial }{\partial
\mu_3}\ln Z(\mu_3,\mu_8,T),
\]
\[
\mu_8 = \frac{2kT}{K}\frac{\partial }{\partial \mu_8}\ln
Z(\mu_3,\mu_8,T).
\]
For the system described by one-site Hamiltonian
\eqref{One-siteHamilt}, the self-consistent relations get the form
\[
\mu_3 =\frac{2\sh \tfrac{(J-\frac{K}{2})
\mu_3}{kT}}{\exp\Bigl\{-\tfrac{\sqrt{3}\,K \mu_8}{2kT}\Bigr\}+
2\ch \tfrac{(J-\frac{K}{2}) \mu_3}{kT}},
\]
\begin{equation}\label{SelfCond}
\mu_8 = \frac{2}{\sqrt{3}} \frac{\ch \tfrac{(J-\frac{K}{2})
\mu_3}{kT}-\exp\Bigl\{-\tfrac{\sqrt{3}\,K
\mu_8}{2kT}\Bigr\}}{\exp\Bigl\{-\tfrac{\sqrt{3}\,K
\mu_8}{2kT}\Bigr\}+ 2\ch \tfrac{(J-\frac{K}{2}) \mu_3}{kT}}.
\end{equation}

Note, that the true averages are always less than their expectation
values calculated from the self-consistent relations. Therefore,
solutions of \eqref{SelfCond} have a qualitative sense only.

Here we analyze Eqs. \eqref{SelfCond} and make comparison with
results described in the literature. The obvious solution
corresponds to the paramagnetic state with $\mu_3\,{=}\,0$ and
$\mu_8\,{=}\,0$; this state is realized at temperatures $kT\,{>}\,
J\,{-}\,K/2$. At the same time, this inequality shows that the
model, oriented to ferromagnetic materials, gives a meaningful
result only in the region $J\,{-}\,K/2\,{<}\,0$, which contains
areas with the ferromagnetic and quadrupole orderings, according to
the well-known phase diagram (Fig.~1) for the bilinear-biquadratic
$s\,{=}\,1$ 1-dimensional spin model~\cite{Buchta}.

\begin{center}
\includegraphics[width=0.33\textwidth]{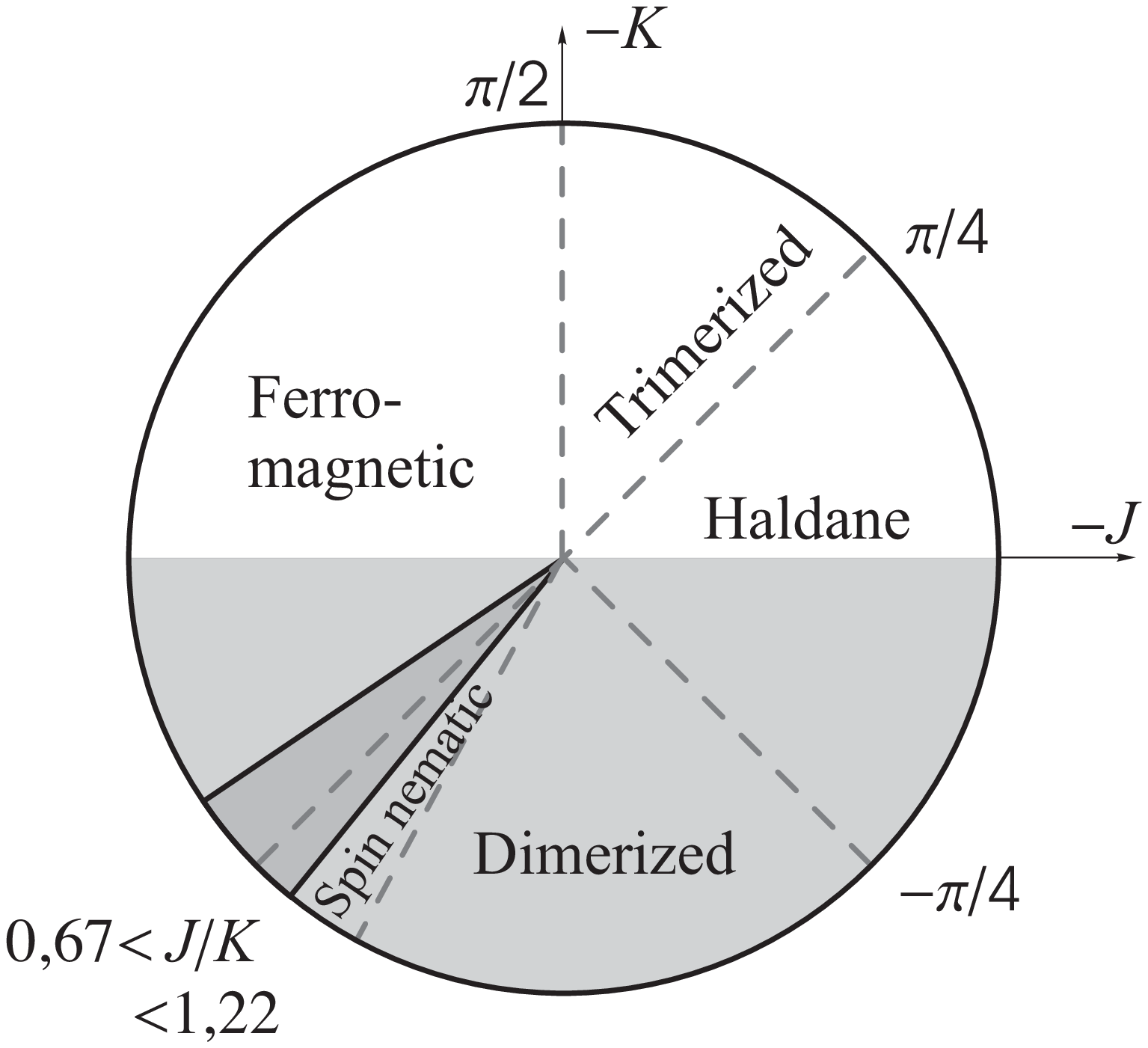}
\end{center}

\vskip-3mm\noindent{\footnotesize Fig.~1. Phase diagram
of a one-dimensional system of spins $s\,{=}\,1$}%
\vskip15pt

In the case of $K\,{<}\,0$, the self-consistent relations have a
unique nontrivial solution, corresponding to the ferromagnetic
ordering, because this solution tends to the boundary values
$\mu_3\,{=}\,1$ and $\mu_8\,{=}\,\frac{1}{\sqrt{3}}$ as temperature
decreases to zero. The critical temperature of transition from a
ferromagnetic state into a paramagnetic one is determined in terms
of the constants $J$ and $K$ as
$T_{\mathrm{c}}\,{=}\,\frac{2}{3k}(J\,{-}\,K/2)$.

In the case of $K\,{>}\,0$ (the light-grey region in Fig.~1), Eqs.
\eqref{SelfCond} have more than one nontrivial solution: two
solutions corresponding to ferromagnetic states with the boundary
values $\mu_3^{(1)} \,{=}\, 1$ and $\mu_3^{(2)} \,{=}\, 2/3$ (and
the corresponding values of $\mu_8$), and two solutions describing
nematic states ($\mu_3\,{=}\,0$) with the boundary values
$\mu_8^{(1)}\,{=}\,\frac{-2}{\sqrt{3}}$ and
$\mu_8^{(2)}\,{=}\,\frac{1}{\sqrt{3}}$. Existence of four ordered
states in ferromagnets is also reported in \cite{Matveev}: they are
a ferromagnetic state with $\mu_3^{(1)}\,{=}\,s$, a quadrupole (or
nematic) state with $\mu_3\,{=}\,0$, $\mu_8^{(1)}\,{=}\, {-}
s(s+1)/\sqrt{3}$, a partially ordered quadrupole state with
$\mu_3\,{=}\,0$, $\mu_8^{(1)}\,{>}\,0$, and a partially ordered
ferromagnetic state with $\mu_3^{(1)}\,{<}\,s$. Partially ordered
states are unstable \cite{Matveev}.

Analyzing the temperature evolution of solutions of \eqref{SelfCond}
as $K\,{>}\,0$,
 $J\,{>}\,0$, we reveal
two critical temperatures, which are solutions of the equation
\[
2\Bigl(\frac{J-K/2}{kT}-1\Bigr)=\exp\Bigl\{ \frac{K}{kT}
\Bigl(1-\frac{3kT}{2(J-K/2)}\Bigr) \Bigr\}.
\]
An obvious solution is $T_{\text{c}1} \,{=}\, \frac{2}{3k}(J-K/2)$.
The other solution $T_{\text{c}2}$ is calculated numerically. In the
region $J\,{>}\,K$, i.e. for ferromagnetic materials, the
temperature $T_{\text{c}2}$ is less than $T_{\text{c}1}$, whereas
the reversed situation takes place for nematics in the region
$J\,{<}\,K$. At smaller critical temperature the solution
$\mu_3^{(2)}$ disappears. Then only the solution $\mu_3^{(1)}$
exists in a certain interval of temperatures. A comparison with
results of the paper \cite{Nauciel} shows that at a higher critical
temperature we have a second order phase transition from a
ferromagnetic state into a paramagnetic one.

A somewhat different behavior is observed in materials with
$J\,{\approx}\, K$, or more precisely $0.67\,{<}\,J/K\,{<}\,1.22$,
i.e. on the boundary between the ferromagnetic and nematic regions
(the dark-grey region in Fig.~1). Disappearing  at a lower critical
temperature, the solution $\mu_3^{(2)}$  arises again at a higher
critical temperature, and grows continuously from zero toward
$\mu_3^{(1)}$. When two solutions coincide at a certain temperature
$T_0$, they disappear by jump. We may conclude that the phase
transition from a ferromagnetic state to a paramagnetic one is a
transition of the first order. This well agrees  with results of the
paper \cite{Nauciel}, where the change of a second order phase
transition into a first order one in the region
$2J/3\,{<}\,K\,{<}\,J$ was considered (for ferromagnetic materials).

Further we consider the case $J\,{=}\,K$, corresponding to the
boundary between the ferromagnetic and quadrupole regions. Moreover,
as $J\,{=}\,K$ the quantum Hamiltonian \eqref{BBH_SU2} and the
Hamiltonian in the mean-field approximation are
$\mathrm{SU}(3)$-invariant. The latter has the form
\[
    \hat{\mathcal{H}}_{\text{MF}}=-\tfrac{1}{2}J \sum_{n}\sum_{a}
    \hat{P}_{n}^{a}\langle
    \hat{P}^{a}\rangle -\tfrac{4}{3}JN =
\]
    \begin{equation}\label{Hamilt3MF} =
    -\tfrac{1}{2}J \sum_{n}\sum_{a} \hat{P}_{n}^{a}\mu_a
    -\tfrac{4}{3}JN.
\end{equation}

\section{Equations of Motion for Large-Scale Fluctuations of the Mean Field}

Now we  return to the quantum $\mathrm{SU}(3)$-invariant spin model
with Hamiltonian \eqref{BBH_SU3}. The Heisenberg evolution equations
for the operators $ \hat{P}_n^a$ have the form
\begin{equation}\label{HeisenbergEq}
i\hbar \frac{d \hat{P}_n^a}{dt}=[\hat{P}_n^a,\hat{\mathcal{H}}].
\end{equation}

We assuming that the system is ordered state, and take an average of
the both sides of Eq. \eqref{HeisenbergEq} over Heisenberg (time
independent) coherent states
\[
|\psi(n)\rangle = \frac{1}{\sqrt{N}} \bigl( c_1 (n) |1\rangle +
c_{-1} (n) |{-}1\rangle + c_0 (n) |0\rangle \bigr),
\]
\[
 |c_1|^2 +
|c_{-1}|^2 + |c_0|^2 =1.
\]
On the other hand, an averaging can be performed by means of the
density matrix (thermodynamical averaging) as $T\,{>}\,0$
\cite{Loktev}. In both cases, we suppose
\begin{equation}\label{ApproxDegenOrb}
\langle \hat{P}^a_n \hat{P}^b_m \rangle \approx \langle \hat{P}^a_n
\rangle \langle \hat{P}^b_m \rangle = \mu_a(n) \mu_b(m),
\end{equation}
i.e. we neglect correlations between fluctuations of the quantum
fields $\hat{P}^a_n$. Then we obtain the following system of
\emph{Hamiltonian equations} for the averages $\mu_a(n)$:
\begin{equation}\label{AvgEqMotion}
\hbar \frac{d \mu_a (n)}{d t} = C_{abc} \mu_b(n) \frac{\partial
\langle\mathcal{H}\rangle}{\partial \mu_c(n)} = \{\mu_a(n),\,
\langle\mathcal{H}\rangle\}.
\end{equation}
Taking \eqref{ApproxDegenOrb} into account we have
$\langle\mathcal{H}\rangle
\,{=}\,\langle\mathcal{H_{\text{MF}}}\rangle$.

In order to investigate large-scale fluctuations of the field
$\mu_a(n)$ we take a two-dimensional continuum instead of the
discrete lattice. Such transition is well known for an
$\mathrm{SU}(2)$-magnet \cite{Herring} and underlies the macroscopic
phenomenological theory of magnetism \cite{Baryahtar}. In the
continuous theory dynamical variables are  densities of averaged
spin and quadrupole moments:
\[
M_a(\bm{x}) = \lim_{S\to 0} \frac{1}{S} \sum_{n\in S} \mu_a(n)
\delta_{\bm{x}, \bm{x}_n} = \sum_{n\in S} \mu_a(n)
\delta(\bm{x}-\bm{x}_n).
\]
Here, $S$ is a `physically' infinitesimal region of the lattice, and
$\delta(\bm{x}{-}\bm{x}_n)$ is the Dirac delta-function, which has
the dimension of reciprocal area. A Poisson bracket on the space of
$\{M_a(\bm{x})\}$ is calculated by the formula
\[
\{M_a(\bm{x}), M_b(\bm{y})\} = C_{abc} M_c(\bm{x})
\delta(\bm{x}-\bm{y}).
\]
In what follows, we deal with the dimensionless field
$\mu_a(\bm{x})\,{=}\,l^2 M_a(\bm{x})$, where $l$ is a distance
between the nearest neighbors of the square lattice.

Considering $(j\,{\pm}\, 1,k)$, $(j,k\,{\pm}\,1)$ as the nearest
neighbors of the site $n\,{=}\,(j,k)$, in Eqs. \eqref{AvgEqMotion}
we perform a transition from the discrete variable $\bm{x}_n$ to a
continuous one $\bm{x}$. Then the field $\{\mu_a(\bm{x})\}$
satisfies the equations
\begin{equation} \label{AvgEqMotionCont}
 \hbar \frac{\partial \mu^a (\bm{x})}{\partial t} =
 \{\mu_a(\bm{x}),\, \mathcal{H}_{{\mathrm{eff}}}\}
= - C_{abc} \mu_b \frac{\delta
\mathcal{H}_{{\mathrm{eff}}}}{\delta \mu_c},
\end{equation}
where
\[
\mathcal{H}_{{\mathrm{eff}}} =J\, \int \sum_a
 \Bigl(\frac{\partial \mu_a}{\partial \bm{x}} \Bigr)^2 \,
 d^2\!\bm{x}.
\]

A suitable representation of the system of Hamilton equations
\eqref{AvgEqMotionCont} is the matrix equation
\begin{equation}\label{MotionEqs}
\frac{\partial \hat{\mu}}{\partial t} = \frac{2Jl^2}{\hbar}
[\hat{\mu},\Delta \hat{\mu}],
\end{equation}
where
\[
 \hat{\mu} = -\frac{i}{2}\sum _a \mu_a(\bm{x}) \hat{P}^a.
\]
Obviously, $\hat{\mu}$ is a Hermitian $3\,{\times}\, 3$ matrix, the
bracket $[\cdot,\cdot]$ denotes a matrix commutator, and
$\Delta\,{=}\,\frac{\partial}{\partial
x^2}\,{+}\,\frac{\partial}{\partial y^2}$ is the Laplace operator
over a two-dimensional space. Equation \eqref{AvgEqMotionCont}
generalizes the Landau--Lifshits equation \cite{Landau} for
isotropic ferromagnets to the case of 8-component mean field
$\mu_a(\bm{x})$. The Landau--Lifshits equation is well-known in the
macroscopic theory of magnetism, and suitable for describing
large-scale excitations in planar magnets and exploring the
ferromagnetic resonance.

\section{Free Energy and Topological Charge}

As mentioned above, a Hamiltonian coincides with a free energy at
constant temperature and volume. Therefore, we use the notion of
free energy in what follows. As shown in Appendix (Section 6) by
means of an algebraic approach, Eq. \eqref{MotionEqs} coincides with
the two-dimensional generalization of Eq. \eqref{EqDegOrbit} on a
degenerate orbit. This equation is a Hamiltonian one, though it is
nonintegrable in the two-dimensional case; and its Hamiltonian can
be used as an effective free energy for the spin system in question,
namely:
\[
\mathcal{F}_1^{{\mathrm{eff}}} = \frac{2}{3h_0} \iint
\bigl(\mu_{a,x}^2 + \mu_{a,y}^2\bigr)\, dxdy.
\]
Obviously, $\mathcal{F}_1^{\text{eff}}$ is a part of the total free
energy of a magnet, and arises from an inhomogeneous distribution of
the average values $\{\mu_a(x)\}$.

The algebraic approach yields one more equation of motion,
associated with a generic orbit. Evidently, this equation can be
obtained from the quantum-mechanical approach, by performing a
relevant averaging that takes correlations into account. Therefore,
we consider another effective free energy
\[
\mathcal{F}_2^{{\mathrm{eff}}} = \frac{1}{2(3f_0^2-h_0^3)} \iint
\bigl(h_0^2 (\mu_{a,x}^2 + \mu_{a,y}^2) -
\]
\[
 - 6f_0
(\mu_{a,x}T_{a,x} + \mu_{a,y}T_{a,y}) + 3h_0 (T_{a,x}^2 +
T_{a,y}^2)\bigr)\, dxdy.
\]
Equations of extremals for the functionals of free energy are the
two-dimensional generalization of Eqs. \eqref{EqDegOrbit} and
\eqref{EqGenericOrbit}.

If equations of constrains determining orbits are solved or, in
other words, orbits are parameterized, then the formulas for the
free energy can be simplified. Orbits of co-adjoint representation
of semisimple compact Lie groups are compact K\"{a}hlerian
manifolds. Therefore, it is suitable to use a complex
parameterization. In order to parameterize a generic orbit, it
requires three complex variables $w_1$, $w_2$, d $w_3$. Explicit
formulas for the parameterization of a generic orbit are the
following:
\end{multicols}
\vspace{-0.2cm} \noindent \rule{0.5pt}{2mm}\rule{85.6mm}{0.5pt}
\[
\mu_1 = \frac{m-\sqrt{3}\,q}{2\sqrt{2}}\cdot
\frac{w_2+w_3+\bar{w}_2+\bar{w}_3}{1+|w_2|^2+|w_3|^2} -
\frac{m}{\sqrt{2}}\frac{(1-w_1)(\bar{w}_3-\bar{w}_1\bar{w}_2)
+(1-\bar{w}_1)(w_3-w_1w_2)}{1+|w_1|^2+|w_3-w_1w_2|^2},
\]
\[
\mu_2= \frac{m-\sqrt{3}\,q}{2i\sqrt{2}}\cdot
\frac{w_3-w_2-\bar{w}_3+\bar{w}_2}{1+|w_2|^2+|w_3|^2} -
\frac{im}{\sqrt{2}}\frac{(1+w_1)(\bar{w}_3-\bar{w}_1\bar{w}_2)
-(1+\bar{w}_1)(w_3-w_1w_2)}{1+|w_1|^2+|w_3-w_1w_2|^2},
\]
\[
\mu_3 =
-\frac{m-\sqrt{3}\,q}{2}\cdot\frac{|w_2|^2-|w_3|^2}{1+|w_2|^2+|w_3|^2}
+\frac{m(1-|w_1|^2)}{1+|w_1|^2+|w_3-w_1w_2|^2},
\]
\[
\mu_4 = \frac{m-\sqrt{3}\,q}{2i}\cdot
\frac{\bar{w}_2w_3-w_2\bar{w}_3}{1+|w_2|^2+|w_3|^2}
+\frac{im(w_1-\bar{w}_1)}{1+|w_1|^2+|w_3-w_1w_2|^2},
\]
\[
\mu_5 = \frac{m-\sqrt{3}\,q}{2\sqrt{2}}\cdot
\frac{w_3-w_2+\bar{w}_3-\bar{w}_2}{1+|w_2|^2+|w_3|^2}-
\frac{m}{\sqrt{2}}\frac{(1+w_1)(\bar{w}_3-\bar{w}_1\bar{w}_2)
+(1+\bar{w}_1)(w_3-w_1w_2)}{1+|w_1|^2+|w_3-w_1w_2|^2},
\]
\[
\mu_6 = \frac{m-\sqrt{3}\,q}{2i\sqrt{2}}\cdot
\frac{w_2+w_3-\bar{w}_2-\bar{w}_3}{1+|w_2|^2+|w_3|^2} +
\frac{im}{\sqrt{2}}\frac{(1-\bar{w}_1)(w_3-w_1w_2)-(1-w_1)(\bar{w}_3-\bar{w}_1\bar{w}_2)}
{1+|w_1|^2+|w_3-w_1w_2|^2},
\]
\[
\mu_7= \frac{m-\sqrt{3}\,q}{2}\cdot
\frac{\bar{w}_2w_3+w_2\bar{w}_3}{1+|w_2|^2+|w_3|^2}
-\frac{m(w_1+\bar{w}_1)}{1+|w_1|^2+|w_3-w_1w_2|^2},
\]
\begin{equation}\label{ParamVar} \mu_8
=-\frac{m-\sqrt{3}\,q}{2\sqrt{3}}\cdot\frac{2-|w_2|^2-|w_3|^2}{1+|w_2|^2+|w_3|^2}
+\frac{m}{\sqrt{3}}\cdot
\frac{1+|w_1|^2-2|w_3-w_1w_2|^2}{1+|w_1|^2+|w_3-w_1w_2|^2}.
\end{equation}
\begin{multicols}{2}
\vspace{-0.2cm} \noindent Here, $m$ and $q$ denote boundary values
of the quantities $\mu_3$ and $\mu_8$, respectively. For a
degenerate orbit, it is sufficient to have two complex variables,
for instance $w_2$ and $w_3$; in this case, we assign $w_1\,{=}\,0$.

After the restriction onto an orbit by formulas \eqref{ParamVar},
the expressions for free energy get the form
\begin{equation}\label{FreeEn}
 \mathcal{F}^{{\mathrm{eff}}} =
\int \sum_{\alpha, \beta} g_{\alpha\bar{\beta}} \left(
\frac{\partial w_{\alpha}}{\partial z} \frac{\partial
\bar{w}_{\beta}}{\partial \bar{z}} +
 \frac{\partial w_{\alpha}}{\partial \bar{z}} \frac{\partial
\bar{w}_{\beta}}{\partial z}\right) dz d\bar{z},
\end{equation}
where $g_{\alpha\bar{\beta}}$ denote components of a metrics on the
orbit, and real coordinates $x$, $y$ on a plane are changed into
complex ones $z$, $\bar{z}$.

While coadjoint orbits are K\"{a}hlerial manifolds they possess
K\"{a}hler potentials, which generate related metrics tensor $g$ and
2-form $h$; their components are calculated by the formulas
\[
g_{\alpha\bar{\beta}} = \frac{\partial^2 \Phi}{\partial w_{\alpha}
\partial \bar{w}_{\beta}},\qquad h_{\alpha\bar{\beta}} = i
\frac{\partial^2 \Phi}{\partial w_{\alpha} \partial
\bar{w}_{\beta}}.
\]
A 2-form gives rise to a topological charge
\[
Q = \frac{1}{4\pi} \int \sum_{\alpha \beta} h_{\alpha\bar{\beta}}
\left( \frac{\partial w_{\alpha}}{\partial z} \frac{\partial
\bar{w}_{\beta}}{\partial \bar{z}} - \frac{\partial
w_{\alpha}}{\partial \bar{z}} \frac{\partial
\bar{w}_{\beta}}{\partial z}\right) dz\wedge d\bar{z},
\]
which means a degree of mapping of a plane into an orbit, realized
by the function $w_1$, $w_2$, $w_3$.

On a degenerate orbit of $\mathrm{SU}(3)$ the function $\Phi_2
\,{=}\, \ln(1\,{+}\,|w_2|^2\,{+}\,|w_3|^2)$ serves as a
K\"{a}hlerian potential, and the metrics tensor from \eqref{FreeEn}
is a K\"{a}hlerian one. Then the formulas for topological charge and
free energy differ only in a sign (`$+$' for a free energy, and
`$-$' for a topological charge), hence,
\[
\mathcal{F}[\xi] \geqslant 4\pi |Q|.
\]
The equality holds if the second term in the brackets vanishes, that
happens if the functions $\{w_{\alpha}\}$ are holomorphic or
antiholomorphic.

Consequently, holomorphic functions form a class of solutions with
quadrupole ordering ($m\,{=}\,0$) that correspond to the minimum of
free energy. The same takes place for antiholomorphic functions.

Now we consider the case of a generic orbit. The cohomology class of
rank 2 for the orbit is two-dimensional, then there exist two basis
2-forms, generated by the K\"{a}hlerian potentials $\Phi_2$, and
$\Phi_1 \,{=}\, \ln(1\,{+}\,|w_1|^2\,{+}\,|w_3\,{-}\,w_1 w_2|^2)$.
As a unique K\"{a}hlerian potential we take the one corresponding to
the Kirillov-Costant-Suriau form
\begin{equation}\label{KahlerianPotential}
\Phi = m\Phi_1 - \frac{m-\sqrt{3}\,q}{2}\,\Phi_2 .
\end{equation}
Generally speaking, the metrics tensor of free energy \eqref{FreeEn}
is not K\"{a}hlerian. However, we can construct a K\"{a}hlerian
metrics of the form
\[
\mathcal{F}_2^{{\mathrm{eff}}} = \frac{1}{2(3f_0^2-h_0^3)} \iint
\Bigl(C_1 (\mu_{a,x}^2 + \mu_{a,y}^2) -
\]
\[
+ C_2 (\mu_{a,x}T_{a,x} + \mu_{a,y}T_{a,y}) + C_3 (T_{a,x}^2 +
T_{a,y}^2)\Bigr)\, dxdy,
\]
by choosing appropriate coefficients $C_1$, $C_2$, $C_3$.

Then all conclusions relative to a degenerate orbit can be extended
to a generic one. That is, the class of solutions with ferromagnetic
ordering that correspond to the minimum of free energy are realized
by holomorphic (or antiholomorphic) functions.

\subsection*{5.1. Large-scale topological excitations}

 Excitations of a state with
quadrupole ordering are described by spatially inhomogeneous
distributions of the 8-component vector field $\mu_a(x)$, living on
a degenerate orbit
\[
\mathcal{O}(\mu_3=0,\mu_8) \simeq
\frac{\mathrm{SU}(3)}{\mathrm{SU}(2)\times \mathrm{U}(1)}.
\]
Mappings of topological charge 2 can be modeled by the holomorphic
functions
\[
w_2(z)=\frac{a_1}{z-z_1},\qquad w_3(z)=\frac{a_2}{z-z_2},
\]
where  $a_1$, $a_2$, $z_1$, and $z_2$ are fixed complex numbers.

The components $\mu_3$ and $\mu_8$ of the mean field, whose boundary
values are called order parameters, are presented in Figs. 2 and 3
(we show a cut along the straight line joining poles of the
functions $w_2(z)$ and $w_3(z)$, the origin of coordinates being at
a middle of the interval $z_1 z_2$). In Figs. 2 and 3, $q$ is a
value of the component $\mu_8$ at an initial point of a degenerate
orbit ($\mu_3\,{=}\,0$).

\begin{center}
\includegraphics[width=0.3\textwidth]{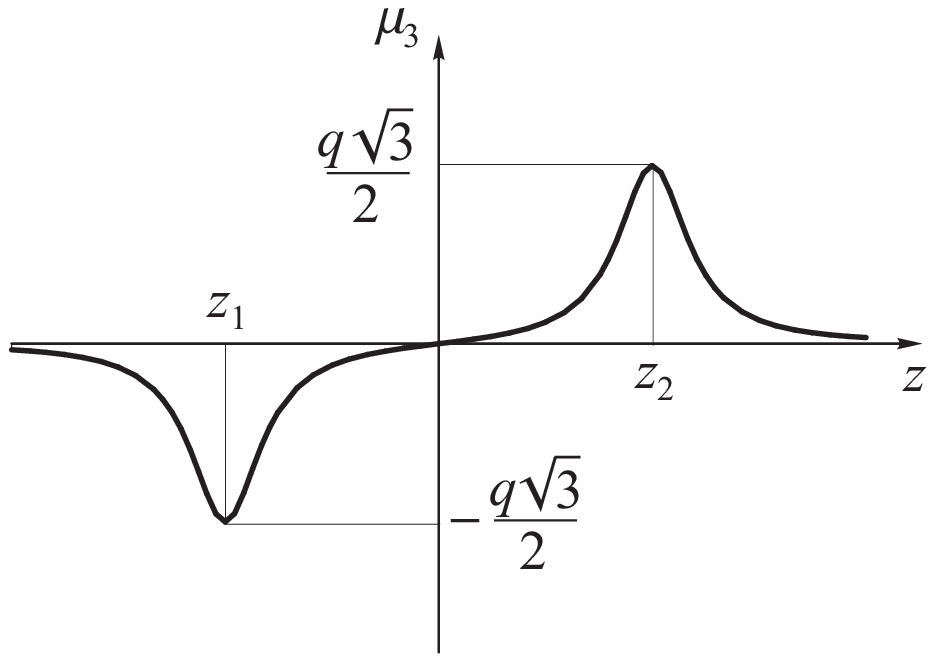}
\end{center}

\vskip-3mm\noindent{\footnotesize Fig.~2. Contour of $\mu_3(z,\bar{z})$, $\mu_3(\infty)\to 0$}%
\vskip15pt

\begin{center}
\includegraphics[width=0.32\textwidth]{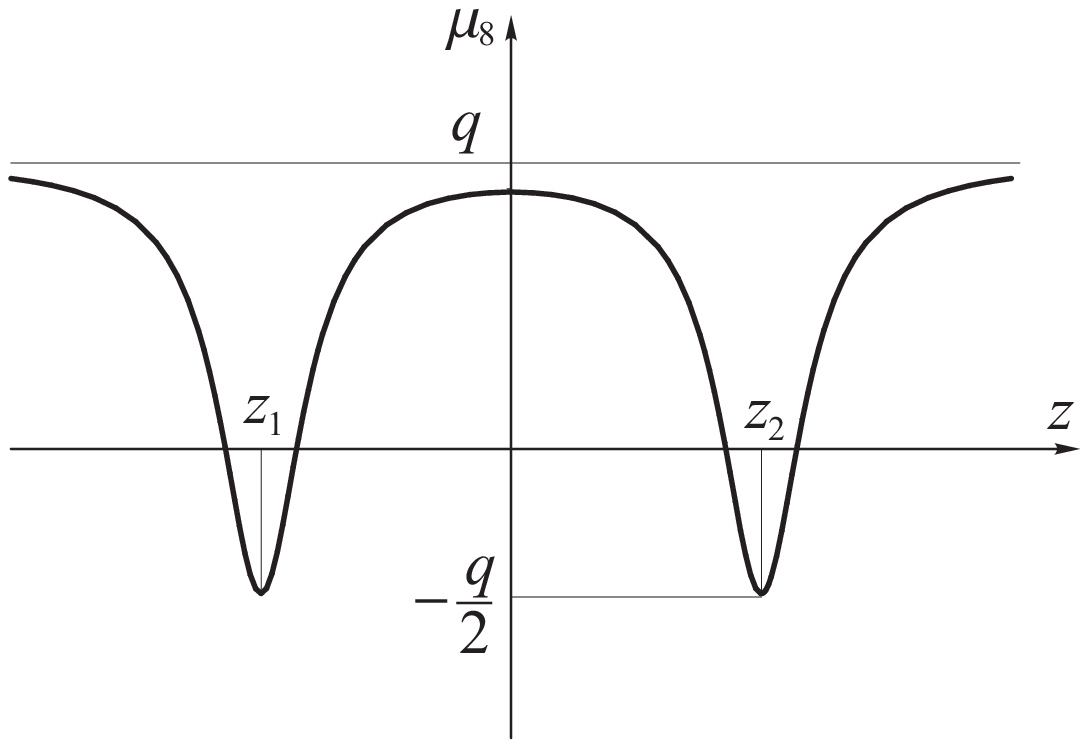}
\end{center}

\vskip-3mm\noindent{\footnotesize Fig.~3. Contour of $\mu_8(z,\bar{z})$, $\mu_8(\infty)\to q$}%
\vskip15pt

These excitations are analogous to Belavin--Polyakov solitons,
well-known for planar Heisenberg ferromagnets (the quantities $a_1$
and $a_2$ define widths of solitons, and the quantities $z_1$, $z_2$
give their positions). It is easy to see that an energy of a
configuration does not depend on a width of soliton, that proves
scale invariance of the energy in the two-dimensional case. Hence,
topological perturbations can have arbitrary large size. Such
instability (an unrestricted increase of soliton width without
pumping of energy) can cause a destruction of the nematic order in
the considered model.

\section{Appendix. Integrability of \boldmath$\mathrm{SU}(3)$-Symmetric Equations of the
Landau--Lifshits Type in a One-Dimensional Space}

It is known that system of equations \eqref{MotionEqs} in the
one-dimensional case is an integrable Hamiltonian system on a
degenerate orbit of the group $\mathrm{SU}(3)$ \cite{Holod}. We
generalize Eq. \eqref{MotionEqs} to the case of a generic orbit. The
algebraic-geometric nature of equations like \eqref{MotionEqs} is
revealed in the frame of so called orbit approach to nonlinear
Hamiltonian systems. Below, we construct integrable Hamiltonian
equations on orbits of the group $\mathrm{SU}(3)$.

Consider polynomials in a complex variable~$\lambda$, whose
coefficients are anti-Hermitian matrices of the algebra
$\mathfrak{su}(3)$. We denote the set of polynomials by
$\widetilde{\mathfrak{g}}_+ \,{\simeq}\,
\mathfrak{su}(3)\,{\otimes}\, \mathcal{P}(\lambda)$, where
$\mathcal{P}(\lambda)$ is a ring of polynomials with the standard
multiplication operation. If $A$, $B\,{\in}\,
\widetilde{\mathfrak{g}}_+$ have the form $A(\lambda)\,{=}\,\sum_n
\hat{A}_n \lambda^n$, $B(\lambda)\,{=}\,\sum_k \hat{B}_k \lambda^k,$
then
\begin{equation}\label{GradOp}
[A,\,B] = \sum_{n,k} [\hat{A}_n,\,\hat{B}_k] \lambda^{n+k}\in
\widetilde{\mathfrak{g}}_+.
\end{equation}
Operation \eqref{GradOp} shows a structure of  graded Lie algebra in
$\widetilde{\mathfrak{g}}_+$. Let $X_a^n \,{=}\, \lambda^n
\hat{X}_a$ be a basis in $\widetilde{\mathfrak{g}}_+$, where
$\hat{X}_a \,{=}\, {-}\frac{i}{2}\hat{\Lambda}_a$,
$a\,{=}\,1,\,2,\,\dots,\, 8$, $\{\hat{\Lambda}_a\}_{a=1}^8$ denote
the Gell-Mann matrices.

In $\widetilde{\mathfrak{g}}_+$ we introduce a bilinear
$\ad$-invariant form
\begin{equation}\label{BilinForm}
\langle A,B \rangle = -2\res \lambda^{-N-2} \Tr A(\lambda)
B(\lambda),
\end{equation}
where $N\,{+}\,1$ is the maximum degree of matrix polynomials $A$
and $B$. Let $\mathcal{M}\,{=}\,(\widetilde{\mathfrak{g}}_+)^{\ast}$
be a space dual to $\widetilde{\mathfrak{g}}_+$ with respect to form
\eqref{BilinForm}. A collection of the linear forms
\[
 \xi(\lambda) = \sum_{n=0}^{N} \sum_{a=1}^8 \xi_a^n \lambda^n \hat{X}_a +
\bigl(\xi_3 \hat{X}_3+ \xi_8 \hat{X}_8\bigr)\lambda^{N+1}
\]
creates a closed $\ad$-invariant subspace $\mathcal{M}^N$ in
$\mathcal{M}$. The coordinates $\xi_a^n$ of $\xi(\lambda)\,{\in}\,
\mathcal{M}^N$ are calculated by the formula
\[
\xi_a^n = \langle \xi(\lambda), X^{-n+N+1}_a \rangle.
\]

In the linear space $\mathcal{M}^N$ with coordinates $\xi_a^n$,
$n\,{=}\,0,\,1,\ldots, N$, we define the Lie--Poisson bracket
\begin{equation}\label{Lie-PoissonBracket}
  \{f_1,f_2\} = \sum_{m,n} \sum_{a,b}^{8}
  W_{ab}^{mn} \frac{\partial f_1}{\partial \xi_a^m}
  \frac{\partial f_2}{\partial \xi_b^n}
\end{equation}
with the Poisson tensor field
\[
W_{ab}^{mn}=\langle \xi(\lambda),
  [X_a^{-m+N+1},X_b^{-n+N+1}]\rangle.
  \]
We also define two $\ad$-invariant functions $I_2(\lambda)$ and
$I_3(\lambda)$ by the formulas
\[
 I_2(\lambda) = -2\Tr
\xi^2(\lambda)= \sum_{a} \xi_a^2(\lambda),
\]
\[
 I_3(\lambda) = -4i\Tr
\xi^3(\lambda) =
d_{abc}\xi_a(\lambda)\xi_b(\lambda)\xi_c(\lambda),
\]
Here, $d_{abc}\,{=}\,{-}2i\Tr (X_a X_b X_c \,{+}\, X_b X_a X_c)$,
and $\xi_a (\lambda)$ is a polynomial \[\xi_a (\lambda) = \xi_a^0 +
\xi_a^1 \lambda + \xi_a^2 \lambda^2 + \cdots + \xi_a^{N+1}
\lambda^{N+1}.\] The invariant functions are also polynomials in the
complex parameter $\lambda$:
\[
 I_2(\lambda) = h_0 +
h_1\lambda + \cdots + h_{2N+2} \lambda^{2N+2},
\]
\[
 I_3(\lambda) = f_0 + f_1\lambda + \cdots+ f_{3N+3}
\lambda^{3N+3}.
\]

It is easy to prove that the coefficients $h_{0}$, $\dots$,
$h_{N+1}$, $f_{0}$, $\dots$ $f_{N+1}$ are annihilators of bracket
\eqref{Lie-PoissonBracket}. Fixing them we obtain the system of
algebraic equations
\begin{equation}\label{OrbitEq}
  h_{n}={\mathrm{const}},\ f_{n}={\mathrm{const}},\
  n=0,\ldots, N+1,
\end{equation}
which give an embedding of the orbit $\mathcal{O}^{N+1}$ of
dimension $6(N+1)$ into the linear space $\mathcal{M}^{N+1}$. The
rest of coefficients $\{h_{N+2}$, $\dots,$ $h_{2N+2}$, $f_{N+2}$,
$\dots,$ $f_{3N+3}\}$ form a pairwise commutative collection of
integrals of motion, which is necessary to integrate the Hamiltonian
system. We are interested in the functions $h_{N+2}$ and $h_{N+3}$
and the corresponding Hamiltonian equations. In particular, the
Hamiltonian $h_{N+2}$ gives rise to so-called stationary equations.
In terms of the coordinates $\xi_a^n$, they are
\begin{equation}\label{StatFlow}
\frac{\partial \xi_a^{n}}{\partial x} = 2f_{abc}\xi_c^0 \xi_b^{n+1},
\end{equation}
where $\{f_{abc}\}$ are antisymmetric structure constants of the
algebra $\mathfrak{su}(3)$ in the basis of Gell-Mann matrices:
\begin{gather*}
[X_a,X_b] = f_{abc}X_c,\\ f_{123}=1,\
f_{458}=f_{786}=\tfrac{\sqrt{3}}{2},\\
f_{147}=f_{165}=f_{246}=f_{257}=f_{345}=f_{376}=\tfrac{1}{2},
\end{gather*}
By $x$ we denote a ``time'' with respect to the Hamiltonian
$h_{N+2}$, which corresponds to evolution equations
\begin{equation}\label{EvolFlow}
\frac{\partial \xi_a^n}{\partial t} = 2f_{abc}(\xi_c^0
\xi_b^{n+2}+\xi_c^1 \xi_b^{n+1}).
\end{equation}

Equations \eqref{StatFlow} and \eqref{EvolFlow} are consistent,
because the corresponding Hamiltonians commute: $\{h_{N+2},\,
h_{N+3}\}\,{=}\,0$. Hence, we can determine evolution
\eqref{EvolFlow} on trajectories of system \eqref{StatFlow}, i.e. we
suppose the dynamical variables $\xi_a^n$ in Eq. \eqref{EvolFlow} to
be dependent on $x$. Combining \eqref{StatFlow} and
\eqref{EvolFlow}, we have
\begin{equation}\label{Nullcurvature}
\frac{\partial \xi_a^0}{\partial t} = 2f_{abc}\xi_c^0 \xi_b^{2}=
\frac{\partial \xi_a^1}{\partial x}.
\end{equation}
The variables $\{\xi_a^1\}$ can be expressed in terms of
$\{\xi_a^0\}$ and $\{\frac{\partial}{\partial x}\xi_a^0\}$, then
\eqref{Nullcurvature} becomes a closed system of partial
differential equations for $\{\xi_a^0\}$. For $\{\xi_a^1\}$ it is
necessary to solve the degenerate system of equations
\begin{equation}\label{StatFlowDegen}
\frac{\partial \xi_a^0}{\partial x} = 2f_{abc}\xi_c^0 \xi_b^{1},
\end{equation}
that becomes possible after restriction to an  orbit
$\mathcal{O}^{N+1}\subset \mathcal{M}^{N+1}$.

\subsection*{6.1. Classification of orbits}

It follows from Eqs. \eqref{OrbitEq} that the orbit
$\mathcal{O}^{N+1}$ in  $\mathcal{M}^{N+1}$ has the structure of a
vector bundle over a co-adjoint orbit of the group $\mathrm{SU}(3)$.
Hence, a classification of the orbits $\mathcal{O}^{N+1}$ is reduced
to that of orbits of  $\mathrm{SU}(3)$.

Since the group $\mathrm{SU}(3)$ is simple, we have
$\mathfrak{g}^{\ast}\,{=}\,\mathfrak{g}$. Therefore, we consider
$\{\xi_a^0\}$ also as coordinates in the algebra
$\mathfrak{g}\simeq\mathfrak{su}(3)$. Then a generic element
$\hat{\xi}\in \mathfrak{su}(3)$ is represented by the matrix
\begin{equation}\label{ElAlgebra}
\widehat{\xi} = -\frac{i}{2}\begin{pmatrix}
\xi^0_3+\frac{1}{\sqrt{3}}\xi^0_8 & \xi_1^0 - i\xi_2^0 & \xi_4^0 - i\xi_5^0 \\
\xi_1^0 + i\xi_2^0 & -\xi_3+\frac{1}{\sqrt{3}}\xi_8& \xi_6^0 - i\xi_7^0 \\
\xi_4^0 + i\xi_5^0&\xi_6^0 + i\xi_7^0&
-\frac{2}{\sqrt{3}}\xi_8\end{pmatrix}.
\end{equation}
Let $\hat{\xi}(0)$ be a fixed element of $\mathfrak{su}(3)$. By
definition, the set
$\mathcal{O}_{\hat{\xi}(0)}\,{=}\,\{g\hat{\xi}(0) g^{-1},\, \forall
g\,{\in}\, \mathrm{SU}(3)\}$ is an orbit of $\mathrm{SU}(3)$ through
the point $\hat{\xi}(0)$. Elements $g'$ of the group
$\mathrm{SU}(3)$ such that $g'\hat{\xi}(0) g'{}^{-1} \,{=}\,
\hat{\xi}(0)$ form the stationary subgroup at the point
$\hat{\xi}(0)$. An orbit $\mathcal{O}_{\hat{\xi}(0)}$, being a
homogeneous space, is a coset space $\mathrm{SU}(3) /
\mathrm{G}'\,{\simeq}\, \mathcal{O}_{\hat{\xi}(0)}$, where
$\mathrm{G}'$ denotes the stationary subgroup.

A maximal commutative subalgebra of a semisimple algebra
$\mathfrak{g}$, which is called Cartan, can be diagonalized. The
Cartan subalgebra of $\mathfrak{su}(3)$ is formed from diagonal
matrices depending on two parameters $\xi_3^0$ and $\xi_8^0$. It~is
well known that a proper transformation puts any element
$\hat{\xi}\,{\in}\,\mathfrak{g}$ into the Cartan subalgebra of
$\mathfrak{g}$. This yields that \emph{each orbit intersects the
Cartan subalgebra at least once}. In fact, there is more than one
intersection point number, more precisely as many as an \emph{order}
of the Weyl group $\mathrm{W(G)}$. We discuss this in what follows.

Nontrivial similarity transformations $\hat{\xi} \,{\to}\,
g\hat{\xi} g^{-1}$ that preserve a subalgebra $\mathfrak{h}$ form a
discrete subgroup $\mathrm{W(G)} \,{\subset}\, \mathrm{G}$, which is
called a Weyl group \cite{Helgason}. An action of the group
$\mathrm{W(G)}$ on the subalgebra
$\mathfrak{h}\,{=}\,\mathfrak{h}^{\ast}$ is generated by reflections
in planes orthogonal to simple roots. The Weyl group of
$\mathrm{SU}(3)$ is generated by two reflections $\sigma_1$ and
$\sigma_2$ in the planes shown in Fig.~4 by dotted lines. The full
Weyl group consists of six elements $\{e,\,\sigma_1,\,\sigma_2,\,
\sigma_1\sigma_2,\, \sigma_2\sigma_1,\,
\sigma_1\sigma_2\sigma_1\simeq \sigma_2\sigma_1\sigma_2\}$, and is
isomorphic to the group of permutations $S_3$, $\ord S_3 \,{=}\,
3!$.

\begin{center}
\includegraphics[width=0.26\textwidth]{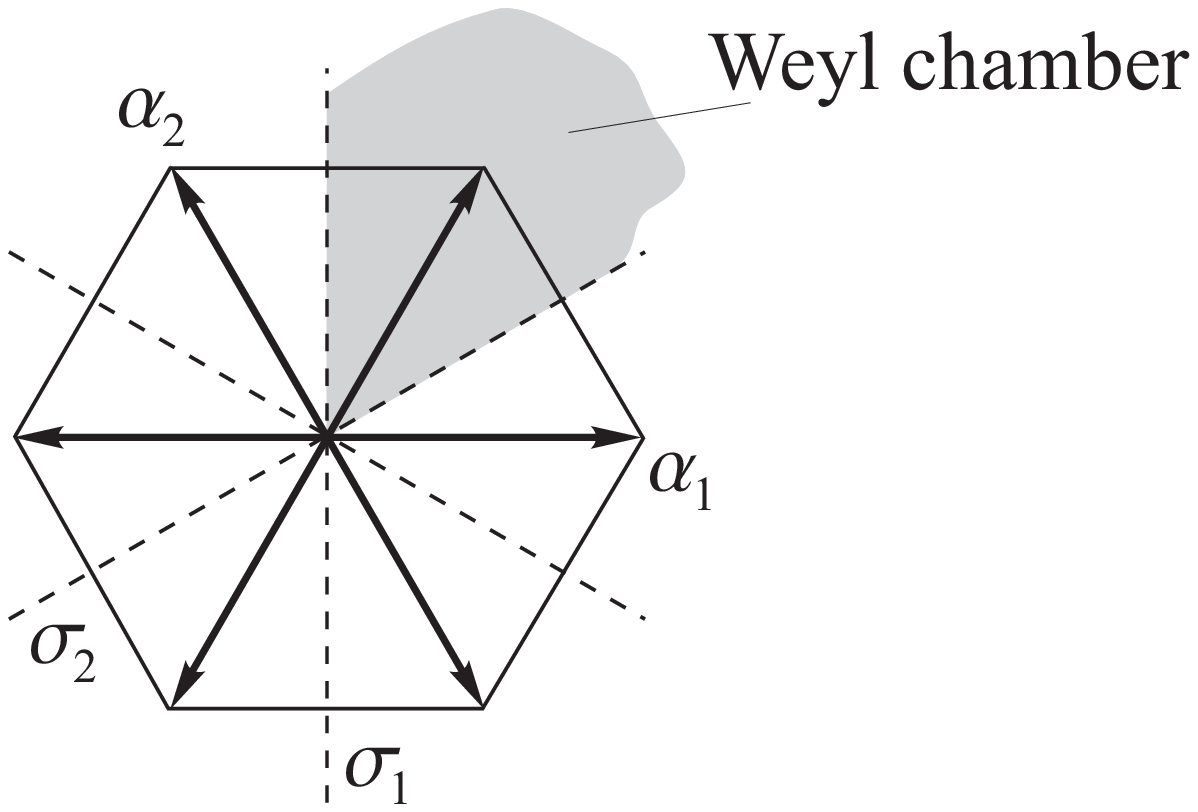}
\end{center}

\vskip-3mm\noindent{\footnotesize Fig.~4. Root diagram of the group $\mathrm{SU}(3)$}%
\vskip15pt

Since a Weyl group acts over a Cartan subalgebra, every point
$\sigma \hat{\xi}(0)\sigma^{-1}$, $\sigma \,{\in}\, \mathrm{W(G)}$,
belongs to an orbit through  $\hat{\xi}(0) \,{\in}\, \mathfrak{h}$.
The group $\mathrm{W(G)}$ acts efficiently (change an element
$\hat{\xi}$ into another, if the element does not belong to
reflection planes). An open domain in a Cartan subalgebra where a
Weyl group acts efficiently, is called \emph{a Weyl chamber} (see
Fig.~4). Elements of different Weyl chambers are adjoint by elements
$\sigma\,{\in}\, \mathrm{W(G)}$, this implies that each orbit
through a point $\hat{\xi}(0)$ of Weyl chamber intersects a Cartan
subalgebra as many times as an order of $\mathrm{W(G)}$. If
$\hat{\xi}(0)$ is an interior point point of a Weyl chamber, we call
the orbit a \emph{generic} one, and call the points $\sigma
\hat{\xi}(0)\sigma^{-1}$, $\forall \sigma \,{\in}\, \mathrm{W(G)}$
 \emph{poles of an orbit}.

For a generic orbit a stationary subgroup $\mathrm{G}'$ coincides
with a maximum torus $\mathrm{T}^r$ for a group $\mathrm{G}$ of
rank~$r$. In the case of group $\mathrm{SU}(3)$, we have
$\mathrm{T}^2 \,{=}\, \mathrm{U}(1)\times \mathrm{U}(1)$. Hence,
generic orbits are coset spaces
$\mathcal{O}_{\text{gen}}\,{\simeq}\, \mathrm{SU}(3) /
\mathrm{U}(1)\,{\times}\, \mathrm{U}(1)$.

If an initial point $\hat{\xi}(0)$ belongs to a wall of a Weyl
chamber (in the case of $\mathrm{SU}(3)$, belongs to one of the
reflection lines), then we deal with a degenerate orbit. In this
case, a stationary subgroup $\mathrm{G}'$ contains a semisimple
subgroup generated by roots orthogonal to an initial point
$\hat{\xi}(0)$. Consider the group $\mathrm{SU}(3)$. If
$\hat{\xi}(0)$ lies the vertical reflection line, $\alpha_1$ and
$-\alpha_1$ are orthogonal to this element. The corresponding
$\mathfrak{sl}(2)$-triple $\{X_{\alpha_1},\, X_{-\alpha_1},\,
H_{\alpha_1}\,{=}\,[X_{\alpha_1},\,X_{-\alpha_1}]\}$ generates a
subgroup $\mathrm{SU}(2)\,{\subset}\, \mathrm{SU}(3)$. Obviously,
the element $\hat{\xi}(0)\,{=}\,\frac{-i}{2\sqrt{3}}\,
\xi^0_8\diag(1,1,-2)$ is invariant under a transformation
$g'\hat{\xi}(0)g'{}^{-1}$, where $g'$ is the unitary matrix
\[
  g' = \begin{pmatrix} \alpha & \beta & 0 \\ -\beta^{\ast} & \alpha^{\ast} & 0\\ 0&0&1  \end{pmatrix}
  \begin{pmatrix} e^{i\varphi/2} & 0 & 0 \\ 0 &  e^{i\varphi/2} & 0\\ 0&0& e^{-i\varphi}
  \end{pmatrix}.
\]
Hence, $g'\in \mathrm{SU}(2)\times \mathrm{U}(1)$, and a degenerate
orbit is a coset space $\mathcal{O}_{\text{deg}}\simeq
\mathrm{SU}(3) / \mathrm{SU}(2)\,{\times}\, \mathrm{U}(1)$.

\subsection*{6.2. Equations on orbits and their Hamiltonians}

Return to construction of integrable systems on orbits of loop
groups, and consider Eq. \eqref{Nullcurvature}. In order to solve
this equation we have to restrict the degenerate system
\eqref{StatFlowDegen} into an orbit and solve.

If $\xi_a^0 \,{\in}\, \mathcal{O}_{\text{deg}}$, then the matrix
$2f_{abc} \xi_c^0$ has rank 4, and its inversion gives a solution
\[
\xi_a^1 = \tfrac{2}{3h_0}\, f_{abc} \xi_b^0  \xi_{c,x}^0 +
\tfrac{h_1}{2h_0}\, \xi_a^0,
\]
where the constants $h_0$, $h_1$ define an orbit by Eqs.
\eqref{OrbitEq}.

If $\xi_a^0 \,{\in}\, \mathcal{O}_{\text{gen}}$, then the matrix
$2f_{abc} \xi_c^0$ has rank 6 and is invertable on a generic orbit.
Then we have
\[
 \xi_a^1 =
\tfrac{1}{2(h_0^3-3f_0^2)}\Bigl(h_0^2 f_{abc} \xi_b^0
 \xi_{c,x}^0 +
\]
\[
 + 3h_0
f_{abc} \eta_b^0  \eta_{c,x}^0 - 6f_0 f_{abc} \xi_b^0
 \eta_{c,x}^0\Bigr) +
\]
\[
  + \tfrac{2f_0 f_1 -
3h_0^2 h_1}{6(f_0^2-h_0^3)}\, \xi_a^0 + \tfrac{3f_0 h_1 - 2h_0
f_1}{6(f_0^2-h_0^3)}\, \eta_a^0,
\]
where $\eta_a^0 \,{=}\, d_{abc} \xi_b^0\xi_c^0$, and the constants
$h_0$, $h_1$, $f_0$, $f_1$ come from Eqs. \eqref{OrbitEq}.

Substituting the obtained expressions in the right-hand side of
\eqref{Nullcurvature}, we get two equations for the functions
$\xi_a(x,t)\equiv \xi_a^0$:
\begin{equation}
\frac{\partial \xi_a}{\partial t} = \tfrac{2}{3h_0}\, f_{abc}
\xi_b \xi_{c,xx} + \tfrac{h_1}{h_0}\, \xi_{a,x},\quad
\xi_a\in\mathcal{O}_{{\mathrm{deg}}}, \label{EqDegOrbit}
\end{equation}
\[
\frac{\partial \xi_a}{\partial t} =
\tfrac{1}{2(h_0^3-3f_0^2)}\Bigl(h_0^2 f_{abc} \xi_b \xi_{c,xx} -
3f_0 f_{abc} \xi_b \eta_{c,xx} +
\]
\[
+ 3h_0 f_{abc} \eta_b \eta_{c,xx} - 3f_0 f_{abc} \eta_b \xi_{c,xx}
\Bigr) +
\]
\begin{equation}
+  \tfrac{2f_0 f_1 - 3h_0^2 h_1}{6(f_0^2-h_0^3)} \,\xi_{a,x} +
\tfrac{3f_0 h_1 - 2h_0 f_1}{6(f_0^2-h_0^3)}\, \eta_{a,x},~~
\phantom{\xi_a}\xi_a\in\mathcal{O}_{{\mathrm{gen}}}.
\label{EqGenericOrbit}
\end{equation}

Let $h_1\,{=}\,0$ in Eq. \eqref{EqDegOrbit}, and replace the
variables $\xi_a$ by $\mu_a$. Then its generalization to the
two-dimensional case gets the form
\[
\frac{\partial \mu_a}{\partial t} = \tfrac{1}{6h_0}\, C_{abc}
\mu_b \Delta \mu_c,
\]
where $\Delta \,{=}\, \frac{\partial^2}{\partial x^2} \,{+}\,
\frac{\partial^2}{\partial y^2}$. Obviously, this equation has the
Hamiltonian
\[
 \mathcal{H}^{{\mathrm{eff}}}
=\tfrac{1}{12h_0} \iint \bigl(\mu_{a,x}^2+\mu_{a,y}^2\bigr)\,dxdy.
\]
It is easy to see that \eqref{EqDegOrbit} coincides with a
one-dimensional analog of \eqref{MotionEqs}. In other words, Eq.
\eqref{MotionEqs} can be considered as a two-dimensional
generalization of the integrable equation \eqref{EqDegOrbit}.

In the same way we treat with Eq. \eqref{EqGenericOrbit}, namely,
replace $\xi_a$ by $\mu_a$ and assign $f_1\,{=}\,h_1\,{=}\,0$.
Generalized to two dimensions, the obtained equations get the form
\[
 \frac{\partial \mu_a}{\partial t} = \tfrac{1}{8(h_0^3-3f_0^2)}
\Bigl(h_0^2 C_{abc} \mu_b \Delta \mu_c -3f_0 C_{abc}
\widetilde{\eta}_b \Delta \mu_c+
\]
\begin{equation}\label{MotionEqsGeneric}
+ 3h_0 C_{abc} \widetilde{\eta}_b \Delta \widetilde{\eta}_c - 3f_0
C_{abc} \mu_b \Delta \widetilde{\eta}_c \Bigr).
\end{equation}
Here, $\widetilde{\eta}_a$ are quadratic forms in $\mu_a$:
$\widetilde{\eta}_a \,{=}\, \widetilde{d}_{abc} \mu_b \mu_c$, where
$\widetilde{d}_{abc} \,{=}\, \frac{1}{4} \Tr (P_a P_b P_c \,{+}\,
P_b P_a P_c)$. Obviously, Eq. \eqref{MotionEqsGeneric} is
Hamiltonian, and give the following effective Hamiltonian
\[
\mathcal{H}^{{\mathrm{eff}}} = \tfrac{1}{16(h_0^3-3f_0^2)} \iint
\bigl(h_0^2 \mu_{a,x}^2 + h_0^2 \mu_{a,y}^2 -
\]
\[
 - 6f_0
\mu_{a,x}\widetilde{\eta}_{a,x}
 - 6f_0 \mu_{a,y}\widetilde{\eta}_{a,y} +
3h_0 \widetilde{\eta}_{a,x}^2
 + 3h_0 \widetilde{\eta}_{a,y}^2 \bigr)\, dxdy.
\]

\section{Conclusions}

In the present paper, we have constructed nonlinear stationary
excitations appearing in the nematic phase of a planar magnet of
spin $s\,{=}\,1$, modeled by a square lattice with a biquadratic
interaction between nearest-neighbor sites. These excitations are
characterized by an integer topological charge, and reveal
themselves as regions with nonzero magnetization and mean quadrupole
moment. Topological excitations in a two-dimensional system (without
taking into account an anisotropy and a demagnetizing field) can
increase unrestrictedly  without pumping of energy. This destroys a
nematic state in the system, according to the Mermin--Wagner theorem
on absence of a long-range order in one- and two-dimensional
systems.

\vskip3mm

This work is partly supported by grant DFFD UkrF16/457-2007 and the
grant of the International Charitable Fund for Renaissance of
Kyiv-Mohyla Academy.

\rezume{%
┬╧╬╨▀─╩╬┬└═▓ ╤╥└═╚  ╥└ ═┼╦▓═▓╔═▓\\ ┬┼╦╚╩╬╠└╤╪╥└┴═▓ ╟┴╙─╞┼══▀\\ ╙
╧╦╬╤╩╬╠╙ ╠└├═┼╥╚╩╙\\ ╟▓ ╤╧▓═╬╠ $s\,{=}\,1$}{▐.╠.~┴хЁэрЎ№ър,
╧.▓.~├юыюф} {─юёы│фцхэю тяюЁ фъютрэ│ ёЄрэш Єр Єюяюыюу│ўэ│
чсєфцхээ  є ътрч│фтютшь│Ёэюьє ьруэхЄшъє, чьюфхы№ютрэюьє ътрфЁрЄэю■
┤ЁрЄъю■ ч│ ёя│эрьш $s\,{=}\,1$ є тєчырї Єр урь│ы№Єюэ│рэюь ч
с│ътрфЁрЄэю■ юсь│ээю■ тчр║ьюф│║■ эрщсышцўшї ёєё│ф│т. ╟ряЁюяюэютрэю
фтр хЇхъЄштэшї урь│ы№Єюэ│рэш фы  юяшёє тхышъюьрё°Єрсэшї чсєфцхэ№ є
ёЄЁюую фтютшь│Ёэюьє тшярфъє. ╬фшэ ч эшї юяшёє║ чсєфцхээ 
ёхЁхфэ№юую яюы  т эхьрЄшўэ│щ Їрч│, │э°шщ~--- є чь│°рэ│щ
ЇхЁюьруэ│Єэю-эхьрЄшўэ│щ Їрч│. ╧юърчрэю, ∙ю хЇхъЄштэ│ урь│ы№Єюэ│рэш
ь│э│ь│чє■Є№ё  эр ъюэЇ│уєЁрЎ│ ї,  ъ│ ьр■Є№ Ї│ъёютрэшщ Єюяюыюу│ўэшщ
чрЁ ф. ╓│ Єюяюыюу│ўэ│ чсєфцхээ  ьюцєЄ№ тшэшърЄш яЁш эхчэрўэшї
ЄхьяхЁрЄєЁрї │ сєЄш яЁшўшэю■ Ёєщэєтрээ  фры№э№юую яюЁ фъє є ёЄЁюую
фтютшь│Ёэ│щ ёшёЄхь│.}

\end{multicols}
\end{document}